# Attribute-Based Semantic Type Detection and Data Quality Assessment

Marcelo Valentim Silva
Curtin University
Perth, WA, Australia
marcelo.valentimsilva@postgrad.curtin.edu.au

Hannes Herrmann
Curtin University
Perth, WA, Australia
hannes.herrmann@curtin.edu.au

Valerie Maxville
Curtin University
Perth, WA, Australia
v.maxville@curtin.edu.au

*Abstract*—The reliance on data-driven decision-making across sectors highlights the critical need for high-quality data; despite advancements, data quality issues persist, significantly impacting business strategies and scientific research. Current data quality methods fail to leverage the semantic richness embedded in words inside attribute labels (or column names/headers in tables) across diverse datasets and domains, leaving a crucial gap in comprehensive data quality evaluation. This research addresses this gap by introducing an innovative methodology centered around Attribute-Based Semantic Type Detection and Attribute-Based Data Quality Assessment. By leveraging semantic information within attribute labels, combined with rule-based analysis and comprehensive Formats and Abbreviations dictionaries, our approach introduces a practical semantic type classification system comprising approximately 23 types, including numerical non-negative, categorical, ID, names, strings, geographical, temporal, and complex formats like URLs, IP addresses, email, and binary values plus several numerical bounded types, such as age and percentage. A comparative analysis with Sherlock, a state-of-the-art Semantic Type Detection system, shows the advantages of our approach in terms of classification robustness and applicability to data quality assessment tasks. Our research focuses on well-known data quality issues and their corresponding data quality dimension violations, grounding our methodology in a robust academic framework. Detailed analysis of fifty distinct datasets from the UCI Machine Learning Repository showcases our method's proficiency in identifying potential data quality issues. Compared to a famous open-source tool like YData Profiling, our method exhibits superior accuracy, detecting 81 missing values across 922 attributes where YData identified only one. By addressing the existing gap through our innovative use of attribute labels for semantic analysis, our method enhances data quality assessment and streamlines the traditionally time-consuming data-cleaning process. This approach has the potential to significantly improve the efficiency and effectiveness of data-driven decision-making across various domains.

*Keywords—Data quality, semantic type detection, missing data*

## I. Introduction

The increasing reliance on data across various sectors underscores the critical importance of ensuring high-quality data before it enters the data cleaning phase. Traditional data quality assessment methods should pay more attention to the semantic richness of attribute information, such as words in attribute labels, names or headers in table columns, or related descriptions. This oversight leads to a significant gap in our ability to identify and understand data quality issues at their source. However, these elements contain valuable semantic information that can be instrumental in identifying data quality issues related to their content, such as ID, name, temporal and geographical information.

TABLE I. EXAMPLE OF A DIRTY DATASET

| Student ID | Last Name | First Name | Age | Country | Humidity | BirthDate |
|---|---|---|---|---|---|---|
| I345343 | white | | 3 | 200 USA | 45 | 3/04/2121 |
| J892932 | Stewart | Ronald | 28 | ? | 70 | 0/1/2010 |
| J892932 | Johnson | Peter | 56 | Australia | -200 | null |

Table I illustrates common data quality problems that may be found by analyzing the formats associated with words in the names of columns/attributes. The first column is the format ID column because it has 'ID' in its name and has wrongly duplicate IDs. Another problem is an improper capitalization found in the Last Name column (format: Name) with the value 'white'. Besides that, the 'First Name' column has a number, 'Age' (Numerical bounded) has a value not in the accepted limits, 'Country' (format: Country) has a '?' in it, Humidity (Numerical Non-Negative) has a negative value and 'BirthDate' (format: Date) shows three unacceptable date values. Our method finds these problems simply by analysing the words in the attributes, their probable format, and their related content. Importantly, it links them with Data Quality Issues and their associated Dimensions as defined in respected research by Visengeriyeva and Abedjan [19].

This research introduces a novel approach to enhance data quality assessment by harnessing the untapped semantic information within attribute labels. Our method identifies and categorises potential data quality issues before traditional data cleaning processes begin by systematically analyzing these labels to derive knowledge about the expected format and data content.

Given the current challenges, we formulated the following research questions:

*A. Research Questions*

This research aims to address the following questions:

*1) Can attribute labels be effectively used for semantic type detection and subsequent data quality assessment?*

*2) How does our attribute-based approach identify data quality issues across diverse datasets?*

*3) What types of data quality issues can our method detect that might be overlooked by traditional data profiling tools?*

*B. Key Contributions*

*1)* A novel approach that leverages semantic information from attribute labels for data quality assessment, complementing existing instance-based methods.

*2)* A practical semantic type classification system comprising 23 types, balancing comprehensiveness with real-world applicability.

*3)* An integrated pipeline that demonstrates significant improvements in detecting data quality issues, including missing values, compared to widely used tools.

*4)* An empirical evaluation of 50 diverse datasets from the UCI Machine Learning Repository, demonstrating the method's effectiveness across various domains and data types.

*5)* A scalable approach capable of efficiently processing large datasets, as demonstrated by our analysis of over 2 million records.

*C. Approach and Significance*

To address these questions, we developed a methodology centered around Attribute-Based Semantic Type Detection and Data Quality Assessment. By leveraging semantic information within attribute labels, combined with rule-based analysis and comprehensive Formats and Abbreviations dictionaries, our approach introduces a practical semantic type classification system comprising approximately 23 types, including numerical non-negative, categorical, ID, names, strings, geographical, temporal, and complex formats like URLs, IP addresses, email, and binary values plus several numerical bounded types, such as age and percentage.

We conduct a comparative analysis with Sherlock, a state-of-the-art semantic type detection system, to validate our approach and demonstrate its effectiveness. This comparison highlights the practical advantages of our method, particularly its focus on common data science types and explicit handling of bounded numerical types, which are crucial for effective data quality assessment.

This shift towards utilizing attribute labels as a preliminary step in data quality assessment promises to streamline data management practices by providing early insights into data quality, potentially reducing the time and resources dedicated to subsequent data cleaning and processing.

This research aims to demonstrate the feasibility and effectiveness of using attribute labels for semantic type detection and data quality assessment. By developing and evaluating a method that interprets the semantic richness of attribute labels, we seek to fill a critical gap in the literature and offer a new pathway for enhancing data quality assessment practices across diverse datasets and domains.

Through a comprehensive evaluation of fifty datasets from the over 600 that exist on the UCI Machine Learning Repository, spanning various domains, our methodology identified 106 data quality issues, including 81 missing values across 922 attributes/columns. This markedly surpasses traditional tools like YData Profiling, which detected only a single instance of missing values. This stark contrast showcases our approach's ability to uncover and address data quality issues effectively and highlights its transformative potential for enhancing data quality assessment across the digital landscape.

The impact of this research extends far beyond academic circles. By enabling more accurate and efficient data quality assessment, our method has the potential to revolutionize data management practices across various industries.

From healthcare, where data quality directly impacts patient outcomes, to finance, where data integrity is crucial for risk assessment and regulatory compliance, our approach offers a powerful tool for enhancing data-driven decision-making.

Moreover, in the era of big data and machine learning, where the quality of input data significantly affects model performance, our method provides a critical first step in ensuring the reliability and effectiveness of AI-driven solutions.

By addressing the existing gap through our innovative use of attribute labels for semantic analysis, our method enhances data quality assessment and streamlines the traditionally time-consuming data-cleaning process. This approach can significantly improve the efficiency and effectiveness of data-driven decision-making across various domains, from business strategy to scientific research, by providing early, accurate insights into data quality issues.

*D. Delimitation*

This research only works where there are words in the label of the attributes in datasets or the names or headers of columns in tables in databases and where content can be easily evaluated against the formats determined.

It evaluates file formats, including .txt, .csv, .data, .xls and .xlsx. Compacted files such as .zip, .tar.gz, and .gz. that contain the previous file formats can also be analysed. When the initial file is a .zip or a .tar.gz file, the user chooses the file to be analysed. Besides that, .gz files contain only one file inside, so it is automatically analysed.

Besides that, it always analyses the first line to check if it is a header or not and excludes the analysis of the first line when it is a header line. It also always evaluates the symbol that separates the data items, allowing ';', ',' and ' ' (blank).

## II. RELATED WORK

Historical and ongoing challenges in data quality have significant implications for both business strategies and scientific research, underscoring the need for robust data quality management practices [7], [12], [21]. A critical inefficiency highlighted in [4] is the disproportionate time spent on data cleaning, which can consume up to 80% of the total analysis time. Our research introduces an innovative solution by automating the detection of data quality issues through the semantic analysis of attribute labels, aiming to streamline the data quality assessment process.

*A. Data quality assessment*

Data quality is crucial for data usability [22]. Data profiling significantly improves data quality assessment findings, enhancing users' understanding of data [14], [15]. Traditional data quality assessments often overlook the semantic richness in attribute labels or column names/ headers, a significant gap our research addresses. Highlighting similar needs, Data X-Ray [23] and reviews of over 660 tools [6] advocate for enhanced automation and clearer methodologies in data profiling. ISO/IEC standards also stress the importance of syntactic, semantic, and pragmatic dimensions in data quality [6], aligning with our focus on attribute labels for a refined evaluation method.

*B. Data quality dimensions and attribute analysis*

Data in datasets, text files, or database tables consist of attributes (columns or fields) describing the data's features or characteristics. These attributes typically have labels, names, or headers, potentially providing semantic cues about the data they represent. Metadata at the column level can further detail the nature of the data, including data type and constraints. Key data quality dimensions critical to attribute analysis include accuracy, completeness, consistency, and timeliness [2, 4]. Literature on leveraging attribute labels for data quality could be more extensive, with [18] being a notable exception. Our research extends beyond .csv formats, enhancing applicability across various file types. By focusing on data quality, our method identifies issues through semantic signals in column names, a strategy that has yet to be widely addressed in current literature.

*C. Data quality issues*

Loshin [13] demonstrated a business rules approach for identifying the underlying causes of poor data quality by transforming declarative data quality rules into actionable code, aligning with our comprehensive strategies across diverse datasets and domains. A study of 22 well-known data quality issues, including missing data and associated data quality dimension violations, has been detailed at [19] and is included in Appendix 1, and a summary is in Table II. The main author's reference provides all appendices discussed in this document [17] and many other products.

TABLE II. SOME DATA QUALITY ISSUES AND THEIR ASSOCIATED DIMENSIONS [19]

| # | Data Quality Issue | Data Quality Dimensions |
|---|---|---|
| 1 | Missing data | Accuracy, Completeness |
| 5 | Extraneous data | Consistency, Uniqueness |
| 6 | Outdated temporal data | Timeliness |
| 9 | Duplicates | Uniqueness |
| 10 | Structural conflicts | Consistency, Uniqueness |
| 15 | Domain violation | Accuracy |
| 17 | Wrong data type | Consistency |

Our research generates alerts similar to those from Ydata Profiling (previously known as Pandas Profiling) [3], such as Missing and Unique Values. Importantly, our alerts are directly mapped to 11 of the 22 Data Quality Issues from the paper [19], which anchors our study into academic knowledge. A detailed comparison of our findings with those from Ydata Profiling illustrates our approach's effectiveness and innovation and is provided later.

*D. Semantic Type Detection*

Semantic type detection is vital for data cleaning, as it identifies data types based on semantic meaning. Several approaches have been developed in recent years, each with its own characteristics:

Hulsebos et al. [10] introduced Sherlock, a deep learning approach to semantic data type detection. Trained on over 680,000 data columns from the VizNet corpus [9], Sherlock employs deep learning to identify 78 distinct semantic types, matching them to column headers from DBPedia [1]. This work has been updated by AdaTyper [11].

AUTOTYPE [24] takes a different approach, synthesizing type detection logic from data types. It detects 84 semantic types with high precision, offering an even more granular classification than Sherlock.

These approaches primarily focus on inspecting data instances to infer semantic types. While they offer comprehensive coverage, the high number of semantic types they identify (78-84) may introduce unnecessary complexity for many practical data quality assessment tasks.

While these systems offer comprehensive coverage of semantic types, the high number of distinctions may introduce unnecessary complexity for many practical data quality assessment tasks.

In contrast to these instance-based approaches, our research initially focuses on analyzing attribute labels/column names before verifying with content analysis. This two-step process allows us to leverage semantic information that might be overlooked by instance-only analysis while still maintaining the ability to verify and refine our classifications based on data content. We propose a more focused set of approximately 23 semantic types, including numerical non-negative, categorical, ID, names, strings, geographical, temporal, and complex formats like URLs, IP addresses, email, and binary values, plus several numerical bounded types. This approach aims to balance comprehensiveness and practicality, maintaining high accuracy while being more directly applicable to common data quality assessment tasks.

To pursue semantic type detection, we employed formats and abbreviations dictionaries. The definition of words to be considered followed a study from the Web Data Commons [16], which analyzed over 90 million tables from the overall Web Table Corpus and obtained a table that connected their work with the cross-domain knowledge base DBpedia [1], resulting in a list with over 1100 of the words commonly found in column names [5]. These are the words found in over a million tables from this study: name, date, title, description, size, and location.

Our formats dictionary ('formats_dictionary.txt' [17]) expands greatly upon this list, associating these common words with probable data formats and containing 1800 words, providing a foundational tool for our attribute analysis.

## III. METHODOLOGY

This section outlines our approach to addressing the research questions, detailing the methods for semantic type detection, data quality assessment across diverse datasets, and identification of data quality issues that traditional tools might overlook.

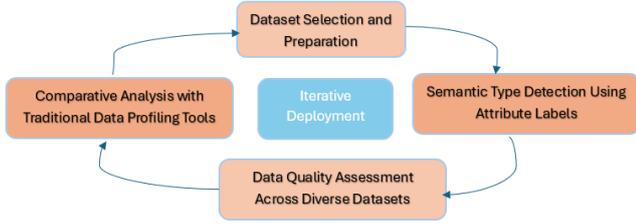

Fig. 1. Methodology of this research

### A. Dataset Selection and Preparation

This section outlines our approach to choosing datasets that effectively test our semantic type detection and data quality assessment methods across various domains and data structures.

- Criteria for ensuring dataset diversity and relevance.
- Process for dataset collection and preprocessing.
- Approach to handling different file formats and data structures.

### B. Semantic Type Detection Using Attribute Labels

This section addresses Research Question 1: *Can attribute labels be effectively used for semantic type detection and subsequent data quality assessment?*

*1) Development of Semantic Analysis Tools*
- Approach for creating and curating the Formats and the Abbreviations Dictionary.

*2) Rule-Based Classification Approach*
- Overview of the classification algorithm for attribute labels/ column names/headers.
- Method for analyzing attribute descriptions.
- Process for handling ambiguities and conflicts in classification.

*3) Semantic Type Classification System*
- Rationale for the chosen semantic types.
- Process for mapping attribute labels to semantic types.
- Approach to handling edge cases and unusual attribute names.

### C. Data Quality Assessment Across Diverse Datasets

This section addresses Research Question 2: *How does our attribute-based approach identify data quality issues across diverse datasets?*

*1) Data Quality Issue Identification Approach*
- Method for validating content against expected formats.
- Process for identifying potential data quality issues.
- Approach to handling format-specific validations.

*2) Data Quality Dimension Categorization*
- Framework for mapping identified issues to data quality dimensions.
- Criteria for assessing issue severity and relevance.

### D. Comparative Analysis with Traditional Data Profiling Tools

This section addresses Research Question 3: *What types of data quality issues can our method detect that might be overlooked by traditional data profiling tools?*

*1) Comparison with YData Profiling*
- Rationale for choosing YData Profiling as a comparison point.
- Process for applying both methods to the selected datasets.

*2) Evaluation Metrics*
- Description of quantitative and qualitative measures
- Justification for chosen metrics in the context of research questions.
- Process for comparing detection capabilities between methods.

## IV. RESULTS

### A. Dataset Selection and Preparation

We successfully collected information from all 622 datasets available in the UCI Machine Learning Repository. This catalogue is a well-known source of research datasets encompassing many domains (described as areas in the catalogue), and some of these datasets have had millions of accesses over time and have been analyzed in dozens of research papers. It was crucial for the generalizability and robustness of our research, ensuring dataset diversity.

We utilized Python libraries, such as Pandas and Beautiful Soup, for web scraping and data gathering, bringing dataset attributes, metadata, and other relevant information necessary for our analysis.

The code used to download information from all 622 datasets and the resulting dataset created are available on GitHub ('UCICatalog-622DataSets.ipynb' and '622_Full_UCI _datasets.csv' [17]).

Two increasing and iterative selections of datasets from the UCI Machine Learning Repository were undertaken to ensure the developed semantic analysis tools were applied to a diverse and representative sample of data. This process aimed to test the tools' applicability across various semantic types and potential data quality issues inherent in different domains.

Our data collection and preprocessing approach was designed to handle a variety of file formats and data structures, as outlined in the Delimitation section in the Introduction.

FIRST DATASET SELECTION

The first selection criteria prioritized ten datasets, ensuring popularity and academic relevance. We sought to cover a wide spectrum of data challenges by choosing datasets from five key areas: Life, Social, Physical, Computer, and Financial.

TABLE III. TEN FIRST DEFINED DATASETS.

| index | name | area |
|---|---|---|
| 52 | Iris | Life |
| 45 | Heart Disease | Life |
| 2 | Adult | Social |
| 107 | Wine | Physical |
| 42 | Glass Identification | Physical |
| 58 | Letter Recognition | Computer |
| 142 | Statlog (German Credit Data) | Financial |
| 92 | Spambase | Computer |
| 103 | Congressional Voting Records | Social |
| 27 | Credit Approval | Financial |

A systematic approach was employed to rank the datasets within each area, combining their web hits and the number of citations each dataset has received in research papers to identify the most significant datasets. This ensured that the selected datasets would provide a robust foundation for our semantic type detection in real-world contexts. For a detailed explanation of the dataset selection criteria and process, please refer to Appendix 2 in the Appendices document [17]). Additionally, Figure 1 in Appendix 2 visually summarizes all the relevant information in these ten datasets, highlighting the diversity and relevance of the chosen data.

This refined selection process identified ten datasets broadly representing potential data quality challenges across various domains ('TenDatasets. xlsx' [17]). See some of the information on these datasets in Table III.

Another important outcome is the file 'AllColumnsFromTenDatasets.xlsx' [17], which is the source for the Semantic Type Detection on the first iteration. It contains information from the ten datasets (index, name, and area) and the content of all the Columns, obtained from the column 'Attribute info' from the '622_Full_UCI_datasets.csv' file.

SECOND DATASET SELECTION

After the initial analysis of the first ten datasets, we expanded our focus to include a second set of forty datasets. The criteria for selecting these datasets are detailed in Appendix 3 [17]. The list of the forty chosen datasets can be found in 'FortyDatasets.xlsx' [17]. Another key document produced from this selection is 'AllColumnsFromForty Datasets.xlsx' [17]. The final number of attributes/ columns obtained from the fifty datasets was 922.

It is important to point out that one of the datasets analyzed, ID 235 - 'Individual household electric power consumption' contained over 2 million records, and the analysis lasted only 20 minutes in a 16 GB RAM notebook. It diminishes any scalability concerns that may exist in this research.

B. Semantic Type Detection Using Attribute Labels

1) Development of Semantic Analysis Tools

The Semantic Analysis Tools developed for this research were the Formats and the Abbreviations Dictionaries. ('formats_ dictionary.txt' and 'abbreviations_dictionary.txt', [17]). The Formats Dictionary was initially assembled with 1000 words associated with specific formats. This was pursued with the help of many conversations with ChatGPT, associated with the Large Language Model GPT 4 from OpenAI. Later, it was expanded with insights from the Web Data Commons and Sherlock's semantic types, resulting in 1800 words/formats. The Abbreviations Dictionary was similarly compiled, translating more than 300 common abbreviations to their full expressions in words of the Formats Dictionary (e.g. pct to percentage).

TABLE IV. FREQUENCY DISTRIBUTION OF THE 1800 WORDS REGARDING THE FORMATS ASSOCIATED IN THE DICTIONARY.

| # | Format | Frequency | Percentage | # | Format | Frequency | Percentage |
|---|---|---|---|---|---|---|---|
| 1 | string | 752 | 41.78% | 18 | country | 1 | 0.06% |
| 2 | categorical | 378 | 21.00% | 19 | day | 1 | 0.06% |
| 3 | numerical | 277 | 15.39% | 20 | hour | 1 | 0.06% |
| 4 | name | 209 | 11.61% | 21 | ID column | 1 | 0.06% |
| 5 | numerical > 0 | 106 | 5.89% | 22 | IP format | 1 | 0.06% |
| 6 | date | 18 | 1.00% | 23 | latitude | 1 | 0.06% |
| 7 | city | 8 | 0.44% | 24 | longitude | 1 | 0.06% |
| 8 | phone | 6 | 0.33% | 25 | model name | 1 | 0.06% |
| 9 | binary | 5 | 0.28% | 26 | month | 1 | 0.06% |
| 10 | datetime | 5 | 0.28% | 27 | normalized | 1 | 0.06% |
| 11 | state | 4 | 0.22% | 28 | numerical between 0 and 360 | 1 | 0.06% |
| 12 | street | 4 | 0.22% | 29 | numerical between 0 and 60 | 1 | 0.06% |
| 13 | postal code | 3 | 0.17% | 30 | percentage | 1 | 0.06% |
| 14 | weekday | 3 | 0.17% | 31 | ph | 1 | 0.06% |
| 15 | E-mail format | 2 | 0.11% | 32 | time | 1 | 0.06% |
| 16 | URL format | 2 | 0.11% | 33 | week | 1 | 0.06% |
| 17 | age | 1 | 0.06% | 34 | year | 1 | 0.06% |
| | | | | | Total | 1800 | 100.00% |

Table IV shows the frequency distribution regarding the associated formats in the Formats dictionary. Observe the 106 words associated with non-negative numbers and the cases in blue that are all numerically bounded, e.g., ' age' and 'day'.

2) Rule-Based Classification Results

The Rule-Based Classification Approach developed in this research was the **Attribute-Based Semantic Type Detection**.

In the initial phase of our data quality assessment, we automated the analysis of attribute labels through a Python notebook titled 'Attribute-BasedSemanticTypeDetection. ipynb' ([17]). This notebook enables a structured approach to identifying potential data formats for each attribute by analyzing target words and abbreviations found in attribute labels and cross-referencing them with our meticulously developed Formats and Abbreviations Dictionaries.

The process is encapsulated in a high-level algorithm, outlined below, and detailed in Appendix 4 [17], which describes the systematic steps taken to extract, transform, and analyze the data from the attribute labels. This Appendix also shows the detailed results associated with it.

---

ALGORITHM 1: Attribute-Based Semantic Type Detection

---

Input: Dataset information file, Formats Dictionary, Abbreviations Dictionary
Output: AnalysedColumns file with semantic types for each column
1: function AttributeBasedSemanticTypeDetection(datasetInfo, formatDict, abbrDict)
2:   columns ← ExtractAndCleanColumns(datasetInfo)

```
3:     for each column in columns do
4:         expandedName ← ReplaceAbbreviations(column.name, abbrDict)
5:         columnFormat ← IdentifyFormat(expandedName, formatDict)
6:         descriptionFormat ← IdentifyFormat(column.description, formatDict)
7:         column.finalFormat ← ResolveFormat(columnFormat, descriptionFormat)
8:     end for
9:     return columns
10: end function

11: function IdentifyFormat(text, formatDict)
12:     keyword ← FindMatchingKeyword(text, formatDict)
13:     return AssociatedFormat(keyword, formatDict)
14: end function

15: function ResolveFormat(columnFormat, descriptionFormat)
16:     // Resolve discrepancies and handle special cases
17:     return finalFormat
18: end function
```

TABLE V. FREQUENCY DISTRIBUTION OF ALL 922 FORMATS.

| ID | FinalFormat | Count | Percentage | ID | FinalFormat | Count | Percentage |
|---|---|---|---|---|---|---|---|
| 1 | numerical | 338 | 36.66 | 16 | longitude | 2 | 0.22 |
| 2 | numerical >= 0 | 243 | 26.36 | 17 | latitude | 2 | 0.22 |
| 3 | categorical | 196 | 21.26 | 18 | datetime | 2 | 0.22 |
| 4 | binary | 76 | 8.24 | 19 | year | 2 | 0.22 |
| 5 | name | 7 | 0.76 | 20 | day | 2 | 0.22 |
| 6 | ID column | 6 | 0.65 | 21 | numerical between 0 and 24 | 2 | 0.22 |
| 7 | NaN | 6 | 0.65 | 22 | time | 2 | 0.22 |
| 8 | age | 6 | 0.65 | 23 | ph | 1 | 0.11 |
| 9 | normalized | 4 | 0.43 | 24 | percentage | 1 | 0.11 |
| 10 | month | 4 | 0.43 | 25 | model name | 1 | 0.11 |
| 11 | date | 4 | 0.43 | 26 | city | 1 | 0.11 |
| 12 | weekday | 3 | 0.33 | 27 | state | 1 | 0.11 |
| 13 | country | 3 | 0.33 | 28 | postal code | 1 | 0.11 |
| 14 | string | 3 | 0.33 | 29 | URL format | 1 | 0.11 |
| 15 | phone | 2 | 0.22 |  | Total | 922 | 100.00 |

*3) Semantic Type Classification System*

Our semantic type classification system was designed to balance comprehensiveness with practicality. We identified 19 distinct formats/types and nine numerical bounded cases such as age or year (in blue in Table V) across the 922 columns analyzed from 50 datasets.

Numerical format: The most common format appears 338 times, about 36.66% of the total.

Numerical >= 0 (Non-negative numbers): The second most common, with 243 instances, making up 26.36%. This allows for quite an interesting analysis, which is not done in other data quality assessments.

Categorical Data: Categorical format was the third most common, identified in 21.26% of columns.

Binary Data: Binary format appeared in 8.24% of columns, indicating a significant presence of data with two values.

Bounded Numerical Types: Several numerically bounded types were identified, such as 'age' and 'normalized' range in [0, 1], allowing for more precise data quality checks.

Rare Formats: Identification formats (e.g., 'name', 'ID column'), geographical formats (e.g., 'country', 'state', 'city'), and temporal formats (e.g., 'date', 'datetime', 'time') were less common but still present.

Unclassified Data (NaN): A small portion (0.65%) of columns could not be classified due to not matching any criteria set in the dictionaries used for semantic type detection.

Other Formats: Three predefined formats ('Street', 'IP Format', and 'Email Format') did not appear in the analyzed datasets, suggesting potential areas for future expansion of our test data.

**Process for mapping attribute labels to semantic types:**

The mapping process involves matching the words in the attribute labels against our Formats Dictionary. We first analyze column names for format identification and, if unsuccessful, examine possible column descriptions. We resolve conflicts between name and description analyses and apply special handling for 'ID' columns and specific patterns. We attempt to find an abbreviation if no format was determined against our Abbreviations Dictionary. Finally, we assign the default 'NaN' classification if no clear format is determined. The corresponding semantic type/format is assigned when a match is found. In cases where multiple matches occur, we apply a priority system based on the specificity of the match and its position in the Formats Dictionary.

*4) Addressing Research Question 1:*

*'Can attribute labels be effectively used for semantic type detection and subsequent data quality assessment?'*

Our results strongly affirm the effectiveness of using attribute labels for semantic type detection. Our approach successfully classified 99.35% of the 922 columns analyzed, with only 0.65% remaining unclassified. The ability to identify 19 distinct formats, including specialized types like non-negative numbers and numerical bounded formats, demonstrates that attribute labels can be leveraged for comprehensive and accurate semantic type detection, laying a solid foundation for subsequent data quality assessment.

TABLE VI. LIST OF 31 SEMANTIC TYPES OF THIS RESEARCH WHERE * IS NUMERICAL BOUNDED

| | | | |
|---|---|---|---|
| age* | email | money | postalcode |
| binary | hour* | name | state |
| categorical | ID | normalized* | street |
| city | IP | numerical | string |
| country | latitude* | numerical>=0 | time |
| date | longitude* | percentage* | weekday |
| datetime | modelname | ph* | year* |
| day* | month | phone | |

TABLE VII. LIST OF 78 SEMANTIC TYPES FROM SHERLOCK [10]

| | | | | |
|---|---|---|---|---|
| Address | Code | Education | Notes | Requirement |
| Affiliate | Collection | Elevation | Operator | Result |
| Affiliation | Command | Family | Order | Sales |
| Age | Company | File size | Organisation | Service |
| Album | Component | Format | Origin | Sex |
| Area | Continent | Gender | Owner | Species |
| Artist | Country | Genre | Person | State |
| Birth date | County | Grades | Plays | Status |
| Birth place | Creator | Industry | Position | Symbol |
| Brand | Credit | ISBN | Product | Team |
| Capacity | Currency | Jockey | Publisher | Team name |
| Category | Day | Language | Range | Type |
| City | Depth | Location | Rank | Weight |
| Class | Description | Manufacturer | Ranking | Year |
| Classification | Director | Name | Region | |
| Club | Duration | Nationality | Religion | |

*5) Comparative analysis with Sherlock*

Tables VI and VII show our and Sherlock's Semantic Types. This comparison is particularly relevant as it allows us to evaluate the effectiveness of our semantic type classification system in the context of state-of-the-art methods in the field.

Our approach differs from Sherlock in several key aspects:

1. **Granularity**: Sherlock offers a classification with 78 types, but our method uses a more coarse-grained approach with ~23 types plus numerical bounded types.

2. **Practicality**: Our focus on common data science types makes our approach more directly applicable to typical data quality assessment tasks.

3. **Handling numerical bounded types**: Unlike Sherlock, we explicitly handle bounded numerical types such as age and percentage, which are crucial for many data quality checks.

4. **Ease of classification**: Our method offers more robust classification with fewer classes, reducing the risk of misclassification.

5. **Semantic Grouping and Extensibility**: We group Sherlock's specific types into broader categories (e.g., 'Artist', 'Creator', 'Director' under the 'name' type in the Formats Dictionary).

Our approach also includes types like URL format, latitude, longitude, ph, phone, and postal code, not directly present in Sherlock's list, making it more comprehensive for certain data quality assessment tasks, especially those involving geospatial data, contact information, and web-related data.

### C. Data Quality Assessment Findings

#### 1) Data Quality Issue Identification Results

The Data Quality Issue Identification Approach developed in this research was the **Attribute-Based Data Quality Assessment**.

During this stage, a Python Jupiter Notebook titled 'Attribute-Based DataQualityAssessment.ipynb' [17] was developed leveraging format identifications from the previous semantic type detection part, critically analyzing the attributes and contents of the selected datasets, aligning with what is in Appendix 1 [17].

Similar to data profiling, this process is encapsulated in a high-level algorithm, described below and detailed in Appendix 5 [17], which describes the systematic steps taken to produce the data quality assessment. This Appendix also shows the detailed results associated with it.

---

ALGORITHM 2: Attribute-Based Data Quality Assessment

---

Input: Dataset files, AnalysedColumns file, Formats Dictionary
Output: Data quality assessment report

```
1: function AttributeBasedDataQualityAssessment(datasetFiles, analysedColumns, formatsDict)
2:     datasets ← LoadDatasets(datasetFiles)
3:     for each dataset in datasets do
4:         columns ← AssignColumnNames(dataset, analysedColumns)
5:         for each column in columns do
6:             format ← DetermineColumnFormat(column, formatsDict)
7:             issues ← ValidateColumn(column, format)
8:             if format is Categorical then
9:                 distribution ← GenerateFrequencyDistribution(column)
10:            end if
11:            ReportIssues(column, issues, distribution)
12:        end for
13:    end for
14:    return CompileQualityReport(datasets)
15: end function
16: function ValidateColumn(column, format)
17:    issues ← ∅
18:    switch format do
19:        case NumericalGreaterEqualZero:
20:            issues ← CheckNumericalGreaterEqualZero(column)
21:        case NumericalBetween:
22:            issues ← CheckNumericalBetween(column)
23:        case ID:
24:            issues ← CheckID(column)
25:        case String:
26:            issues ← CheckStringContent(column)
27:        case Categorical:
28:            issues ← CheckCategorical(column)
29:        case Date, DateTime, Time:
30:            issues ← CheckTemporalFormat(column, format)
31:        case GeographicalFormat:
32:            issues ← CheckGeographicalFormat(column, format)
33:        case SpecialFormat:
34:            issues ← CheckSpecialFormat(column, format)
35:    end switch
36:    return issues
37: end function
38: function CheckDataQualityIssues(column, issues)
39:    for each issue in DataQualityIssueTypes do
40:        if IssuePresent(column, issue) then
41:            issues ← issues ∪ {issue}
42:        end if
43:    end for
44:    return issues
45: end function
```

#### 2) Data Quality Issues and Dimensions Analysis

Below are quantitative measures obtained from the analysis of 106 columns (11,5% from the 922 columns in total) with at least one Data Quality Issue found:

*a) Data Quality Issue (DQI) frequency distribution:*

- Missing Data: 81 instances
- Domain Violation: 7 instances
- Wrong Data Type: 4 instances
- Extraneous Data: 3 instances
- Duplicates: 3 instances
- Uniqueness Violation: 3 instances
- Structural Conflicts: 3 instances
- Non-String Data Type: 2 instances

*b) Data Quality Dimension frequency distribution:*

- Completeness: 81 instances
- Accuracy: 7 instances
- Consistency: 6 instances
- Consistency, Uniqueness: 6 instances
- Uniqueness: 6 instances

#### 3) Key Findings and Missing Value Analysis

Our analysis revealed several significant findings:

- **Missing data identification**: A key result was the precise identification of missing data markers, notably '?', across 14 of the 18 datasets, with Data Quality Issues found in the 50 datasets analyzed (e.g., datasets 45, 2, 103, 27 and ten others). The symbol '?' was found in 78 cases. If we add the two cases where '' (empty) was found and one 'NA', we arrive at 81 columns where Missing Values were found, which means 76,4% of the 106 Data Quality Issues found.

- **Format-specific errors**: Our methodology identified format-specific errors, including implausible negative values (e.g., humidity levels in dataset 360 - Air quality).
- **Geographical and temporal format-specific analysis**: Detailed geographical (such as formats city, street, state, and country in dataset 225 - Restaurant & consumer data) and temporal analyses revealed capitalization errors in geographical columns, and validating complex formats such as URLs, postal codes, and binary formats.
- **Advanced error detection**: These were showcased by identifying non-numerical values in numerical columns (e.g., 'InvoiceNo' and 'StockCode' in dataset 352 - Online Retail) and uncovering instances with over 4000 categories in the 'Description' column in the same dataset 352.

Among the results presented are the two 'Discoveries' documents that exhibit all the outputs provided for each set of datasets analyzed (10 + 40 datasets) [17]. Two spreadsheets were created to analyze the output from all Discoveries. The first one summarizes all Data Quality Issues associated with their Data Quality Dimensions and explains situations when the formats were tested with created Bad Data. All 23 different formats have been tested with many different Bad Data, which are presented in Appendix 6 [17].

Besides that, the second spreadsheet, with the summary of all Data Quality Issues associated with their Data Quality Dimensions and explanations in the 18 datasets among the 50 analyzed that had Data Quality Issues, is presented in Appendix 7 [17]. The spreadsheet 'SummaryofDiscoveries.xlsx' [17] contains all the results.

4) *Addressing Research Question 2:*
*'How does our attribute-based approach identify data quality issues across diverse datasets?'*

Our attribute-based approach has demonstrated robust performance in identifying data quality issues across diverse datasets. We identified 106 columns with data quality issues out of 922 analyzed (11.5%), spanning 18 of 50 datasets (36%).

The approach was particularly effective in detecting missing values, with 76.4% of the identified issues relating to missing data. This performance across various dataset types and domains underscores the versatility and effectiveness of our attribute-based approach in identifying a wide range of data quality issues.

D. *Comparative Analysis with Traditional Data Profiling Tools*

1) *Comparison with YData Profiling*

To validate our approach and demonstrate its effectiveness, we conducted comparative analyses with YData Profiling (previously known as Pandas Profiling), one of the best-performing Data Profiling open-source libraries in Python [8] downloaded over 50,000 times per month and used by millions of data analysts worldwide [19]. This comparison was structured around applying YData Profiling and our research methodologies to the same set of fifty datasets and 922 attributes/columns. The two 'Discoveries' documents were improved with each YData Alerts Analysis in all fifty datasets' explanations.

One example of each alert found from all YData analyses on the fifty datasets evaluated in this research is shown below:

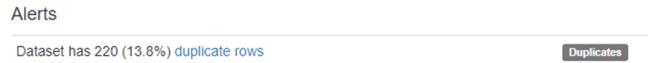

Observe that they analyse Duplicates, but their duplicates relate to the whole record, with all columns. Our research analyses duplicate values only at the column level themselves and only for ID formatted columns. So, we measure differently and, unfortunately, cannot compare.

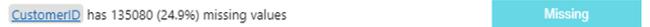

YData checks for Missing values, as seen in the only case described above that they found (related to '' or empty value on dataset 352 - Online Retail). We also check for that but in a broader way. We found not only that instance but also 78 '?'s, two empty values and one 'NA', which also qualified the column as Missing.

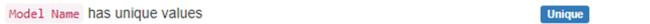

YData measures Uniqueness when there is no repetition in all their data for a column. We only measure this kind of uniqueness when we are evaluating probable ID columns that need to have Unique values so that they can be considered apt to become a Primary Key.

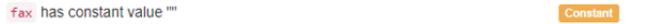

YData measures when the content for a column has only one value, which can be considered a Constant. We do not measure that because it is not in the Data Quality Issues table in Appendix 1 [17].

Below are other five alert analyses YData produces. Our research does not evaluate them as they do not relate to anything according to the Data Quality Issues exhibited in Appendix 1.

TABLE VIII. SOME ALERTS FROM YDATA PROFILING

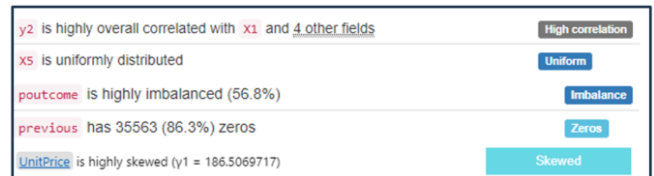

And finally, to our surprise, YData has been producing an 'Unsupported' output for some columns being analyzed. We observed 11 different cases where YData could not evaluate the column. In all these cases, we successfully evaluated the columns and identified data quality issues in almost all of them.

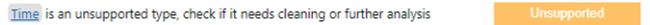

**Key findings and outcomes**

Our analysis revealed several critical areas where YData Profiling Alerts fall short in addressing the nuanced needs of comprehensive data quality assessment:

TABLE IX.   QUALITATIVE RESULTS OF COMPARISON WITH YDATA PROFILING

| Situation | Our research | Ydata Profiling Alerts |
|---|---|---|
| Content '?' | 77 | 0 |
| Missing Values | 81 | 1 |
| Negative Values | 2 (e.g. Humidity) | 0 |
| Geographical and Temporal data | 5 | 0 |
| Non-String and Non-Numerical data | 4 | 0 |
| Time and others | 11 Analyzed and validated | 11 Unsupported |

- **Data type issues**: YData Profiling always missed the content '?' in datasets where they were prevalent, indicating a gap in its semantic analysis capabilities. It only found one missing case where the content was empty.
- **Domain-specific error detection**: The tool did not identify negative values as problematic in contexts where they are implausible, such as humidity levels, highlighting a lack of format-specific checks.
- **Unsupported data types**: YData Profiling showed limitations in supporting and analyzing certain column titles and data formats, such as 'Time', which our method not only supported but validated for expected ranges.
- **Temporal and Geographical data analysis**: We provided detailed analysis for temporal and geographical data, as well as specific format checks like URL, Postal Code, and Binary formats, while YData Profiling did not.

*2) Addressing Research Question 3:*
*'What types of data quality issues can our method detect that might be overlooked by traditional data profiling tools?'*

Our comparative analysis reveals that our method can detect several types of data quality issues often overlooked by traditional tools like YData Profiling:

1. Identification of '?' as missing values, which YData Profiling consistently missed.
2. Detection of domain-specific errors, such as negative values in non-negative fields (e.g., humidity levels).
3. Ability to analyze and validate a broader range of data types, including temporal and geographical data.
4. Capacity to handle data formats that YData Profiling marked as 'unsupported'.

These findings clearly demonstrate the enhanced detection capabilities of our attribute-based approach compared to traditional data profiling tools, addressing complex data quality issues that might otherwise go unnoticed.

## V.   CONCLUSION

This research introduced the Attribute-Based Semantic Type Detection and Data Quality Assessment approach, addressing the critical need for improved data quality assessment methods. By leveraging semantic information embedded in attribute labels, our method offers a novel solution to persistent challenges in data management.

*A. Key Findings and Significance:*

1. Effectiveness of attribute labels: Our approach successfully classified 99.35% of 922 columns analyzed across 50 datasets, demonstrating the power of semantic analysis in data quality assessment.

2. Superior detection of data quality issues: We identified 81 instances of missing values, compared to only one detected by YData Profiling, showcasing our method's enhanced capabilities.

3. Versatility across data types: Our method effectively handled various data types, including numerical, categorical, temporal, and geographical data, highlighting its broad applicability.

*B. Impact and Implications:*

The impact of our approach extends far beyond academic circles, with significant implications for data management practices across various domains:

1. Large-scale data quality management: Our method's ability to quickly assess data quality across thousands or millions of datasets makes it particularly valuable for organizations dealing with big data, enabling proactive data management through real-time quality assessments.

2. Enhanced machine learning performance: By identifying and addressing data quality issues early, our approach can significantly improve model accuracy in critical fields such as predictive maintenance, fraud detection, and personalized medicine.

3. Regulatory compliance: In industries with strict data quality requirements, such as finance and healthcare, our method can aid organizations in complying with regulations by providing comprehensive data quality assessments and audit trails.

4. Cost and time savings: By streamlining the data cleaning process, our method can lead to substantial savings in data preparation, which typically accounts for a significant portion of data scientists' work.

5. Improved decision-making: Our approach gives decision-makers greater confidence in their data-driven strategies by providing more accurate and comprehensive data quality assessments.

*C. Real-world applications of our method could include:*

1. In healthcare: Identifying inconsistencies in patient data formats across different hospital systems, potentially preventing medication errors or improving the accuracy of epidemiological studies.

2. In finance: Enhancing fraud detection systems by detecting subtle data quality issues in transaction data, potentially saving financial institutions millions in fraudulent transactions.

3. In e-commerce: Improving recommendation systems and search functionality by identifying quality issues in product attribute data, enhancing customer experience and potentially increasing sales.

*D. Limitations:*

While our approach shows promising results, it's important to note that it was tested on a specific set of datasets from the UCI Machine Learning Repository. Further testing across a wider range of data sources and domains will be beneficial in establishing its generalizability fully.

In conclusion, our Attribute-Based Semantic Type Detection and Data Quality Assessment approach significantly advances data quality management. By leveraging the semantic richness of attribute labels, we have demonstrated a more comprehensive and nuanced method for identifying data quality issues. As we continue to develop and refine this approach, we anticipate its growing impact on improving data quality across various industries and applications, ultimately leading to more reliable and efficient data-driven processes.

## VI. FUTURE WORK

Our research is set to evolve further, focusing on key advancements:

- Machine Learning and Large Language Models Integration: We plan to incorporate machine learning for more efficient, automated semantic type detection, reducing manual effort and improving adaptability to diverse data, following what was done in the Sherlock paper. We also intend to use Large Language Models API to improve the maintenance of dictionaries.
- Expanding Dataset Analysis: Broadening our analysis beyond the UCI Repository to include 50 additional datasets from varied sources will enhance our framework's generalizability and refine its detection capabilities. We also plan to analyze database tables from an Open-Source Database - db.relational-data.org.
- Expanding Comparison with other Data Profiling Alert systems: Besides YData Profiling, we intend to analyze other Open-Source libraries that provide the same alert results, such as DataPrep.EDA and Autoviz.
- Adhering to ISO and Industry Standards: Aligning with international data quality standards like ISO and Industry Standards such as HL7, SNOMED CT (Health), FIBO and XBRL (Finance), and GS1 and EDI (Commerce) will ensure our methodology meets global data quality benchmarks, increasing its applicability and credibility.

These directions aim to refine our approach, ensuring it remains at the forefront of data quality assessment through innovative techniques and adherence to global standards.